\documentclass[preprint2]{aastex}
\usepackage{amssymb}
\usepackage{amsmath}

\usepackage{graphics}

\newcommand{\be}{\begin{equation}}
\newcommand{\ee}{\end{equation}}
\newcommand{\bd}{\begin{displaymath}}
\newcommand{\ed}{\end{displaymath}}

\slugcomment{submitted to ApJ, \today}

\shorttitle{Thermal X-ray iron line from GX 339-4}
\shortauthors{Y-D Xu }

\begin{document}

%\title{On the presence of outflow/wind in advection dominated accretion
%flow: modeling thermal X-ray iron line emission from galactic binary
%GX 339-4 in the off state }
\title{Thermal X-ray iron line emission from the advection dominated accretion flow
in the galactic binary GX 339-4}

%\author{Ya-Di Xu$^{1}$, Xinwu Cao$^{2}$}
%\affil{$^{1}$Physics Department, Shanghai Jiao Tong University, 800
%Dongchuan Road, Shanghai 200240, China}

%\affil{$^{2}$Key Laboratory for Research in Galaxies and Cosmology,
%Shanghai Astronomical Observatory, Chinese Academy of Sciences, 80
%Nandan Road, Shanghai 200030, China\\
%Email: ydxu@sjtu.edu.cn, cxw@shao.ac.cn}

\author{Ya-Di Xu}
\affil{Physics Department, Shanghai Jiao Tong University, 800
Dongchuan Road, Shanghai 200240, China; Email: ydxu@sjtu.edu.cn}

\clearpage

\begin{abstract}

We explore thermal X-ray iron line emission from the galactic X-ray
binary GX 339-4 in the off state, using the models of the advection
dominated accretion flow (ADAF) without or with outflows. The
equivalent widths of hydrogen-like and helium-like thermal iron
lines are calculated with different model parameters including
viscosity parameter $\alpha$, mass accretion rate at the outer
radius of the ADAF $\dot{m}_{\rm out}$ and outflow strength
parameter $p$. Our calculations show that the equivalent widths of
thermal iron lines emitted from the pure ADAF, i.e.,  the ADAF
without outflows, should be very small, assuming a solar metallicity
for the accreting gas in the accretion flow. Strong thermal iron
lines are expected to be emitted from the ADAF with relatively
strong outflows. {For reasonable choice of parameters, the total
equivalent width of the He-like and H-like thermal iron lines
reaches to $\gtrsim 500$ eV for accreting gas with solar
metallicity. The observation of strong thermal X-ray lines from GX
339-4 at the off state may give a clue to the accretion mode of the
source and provide evidence for the presence of outflows/winds in
the accretion flow around the black hole in GX 339-4. It is found
that the values of $\dot{m}_{\rm out}$ and $p$ are degenerate, i.e.,
the observed X-ray continuum spectrum can be fairly well reproduced
with different sets of the parameters $\dot{m}_{\rm out}$ and $p$.
Such {degeneracy} can be broken when the thermal X-ray line emission
data is available. {We also compare our results with those in the
previous similar work.}}

\end{abstract}

%\clearpage

\section{Introduction}

The galactic X-ray binary (XRB) GX 339-4, one of the earliest
proposed black hole candidates, has been attracted much attention
after discovered in 1971 by the OSO-7 satellite \citep{m73}. Based
on the long-term broadband energy spectral observations, from the
radio, infrared and optical, to the X-rays, GX 339-4 has been
extensively studied on various issues
\citep{f99,c00,k00,n02,m04,m06,d08,t08}. Its spectral and temporal
X-ray properties show a wide variety of canonical XRB states,
transiting from off/{quiescent} state, low/hard state, intermediate
state, high/soft state, and very high state, but it spends most of
its time in the low/hard state. A common argument is that the
spectral states transitions are related to the changes of the mass
accretion rate and the geometry of the accretion flow, though the
physical mechanism of the changes are still quite uncertain
\citep{nmq98,d01}.

Many different models were suggested to explain the observations of
the different spectral states \citep*[see][and references
therein]{nmq98,d01}. The disk+corona models with different
geometries, such as, an accretion disk corona (ADC) above or
surrounded by a cold thin disk, or other similar geometries
\citep*[see][for details and references therein]{n02}, are used to
interpret the observational features of the accretion flows around
the black hole in different states. The observed X-ray spectra in
different spectral states can be explained by the Comptonization (
e.g., soft photons from the cold thin accretion disk are Comptonized
in a hot corona) with different model parameters \citep{n02,c03}.
\citet{n96} proposed that these different spectral states can be
explained by outer thin disk plus inner advection dominated
accretion flow (ADAF) models with varying accretion rate and
transition radius \citep{e97,e98}. In this scenario, the transition
radius increases with decreasing mass accretion rate $\dot{m}$
($\dot{m}=\dot{M}/\dot{M}_{\rm Edd}$). {There is a critical
accretion rate $\dot{m}_{\rm crit}$, above which the ADAF is
suppressed and a standard thin disk is present in the inner region
of the accretion flow.} \citet{m04,m06} suggested that a standard
thin disk may extend at or near to the innermost stable circular
orbit (ISCO), at least in bright phases of the low/hard state, after
{fitting} the spectra of GX 339-4 observed by Chandra during its
decline of 2002-2003 outburst and by XMM-Newton during its 2004
outburst with a cool thin disk component and a relativistic Fe K
line model.  {It is shown} that the disk reflection model can also
fit the spectra in the hard state when the total 1-100 keV
luminosity even lower than $0.023L_{\rm Edd}$ \citep{t08}. Moreover,
the presence of jet/outflow in the accretion flows {has been
explored} to explain the radio emission and the radio/X-ray
correlations in the low/hard state \citep{f99,c00,m03,corbel03}.

Fluorescent Fe K$\alpha$ line emission is a very powerful trace of
the inner edge ($R_{\rm in}$) of the thin accretion disk around the
black holes. \citet{t09} used Suzaku and RXTE observations of {GX
339-4} during 2008 to study the iron line profile at low luminosity
state when the total 1-100 keV luminosity is $0.0014 L_{\rm Edd}$,
and showed that the inner edge of the thin disk recedes and the
truncated accretion disk is present around a black hole at low
luminosity.  If the ADAF exists in the inner region of the accretion
flow as suggested by \citet{n96}, thermal X-ray iron lines may also
emitted from the hot plasma in the ADAF surrounding the black hole
at very low luminosity \citep{n99,p00,xu06,xu09}. {{It is shown}
that the average 0.4-12 keV X-ray flux that measured by RXTE between
2001 March and 2002 February and also on 2003 September 29 is $\sim
5.4\times 10^{-12}{\rm erg~cm^{-2}~s^{-1}}$, when GX 339-4 was in
{the quiescent state} \citep{t09}. This average flux corresponds to
a luminosity of $L_{\rm X}\simeq 4.2\times 10^{34}{\rm
erg~s^{-1}}\sim 3.2\times 10^{-5}L_{\rm Edd}$, if we adopt values of
$m=10$ ($m=M/M_{\odot}$) and $d=8$ kpc for GX 339-4. Chandra even
detected a 0.4-11 keV flux of $3.8\times 10^{-13}{\rm
erg~cm^{-2}~s^{-1}}$ {on 2003 September 29} \citep{g03}, which is
more than an order of magnitude lower than the {average quiescent
value above}. In such very low luminosity quiescent states, the
accretion rate is much lower than the critical accretion rate
$\dot{m}_{\rm crit}$, the outer thin disk may be suppressed, and the
temperature of the plasma in the outer region of the ADAF may be low
to $T_{\rm e}\sim 10^{7-8}$K, so the He-like (Fe XXV) and H-like (Fe
XXVI) thermal X-ray iron lines centered at energy $\sim6.7$ keV and
$\sim6.97$ keV, respectively, can be produced in this region of the
ADAF.} {\citet{n99} calculated {the thermal X-ray line emission}
including He-like and H-like iron lines from another X-ray {binary,
V404 Cyg,} in quiesecence with the ADAF models with and without
winds, and concluded that an ADAF with a large outer radius would
produce {reasonably strong} thermal iron lines if winds are present
in the source. However, for the ADAF models with winds, the
dynamical structure of the ADAFs are solved based on some simplified
assumptions in \citet{n99}. In this work, we will use the improved
numerical approach to derive the global two-temperature structures
of the ADAF models by solving the full relativistic hydrodynamical
{equations with the} modeling techniques suggested by \citet{m00},
in which all the radiation processes are included. }

{In this work, we model the thermal X-ray iron line emission from
the galactic X-ray binary GX 339-4 in the off state, using the
advection dominated accretion flow {(ADAF) models with or without
outflows} \citep{b99}. The models are constrained by the Chandra's
observed X-ray continuum, the power-law photon index of $2\pm0.2$
and the integrated 0.4-11 keV flux of $3.8\times 10^{-13}{\rm
erg~cm^{-2}~s^{-1}}$ \citep{g03}. }We describe the calculation of
dynamical structures of the ADAF in \S 2. The calculation of the
equivalent widths of the thermal X-ray iron {line emission is}
described in \S 3. {The results are shown} in \S 4. In \S 5., we
discuss the results and the physical implications of the results,
{and compare our results of V404 Cyg with those in \citet{n99}.}

\section{Dynamical structures of the ADAF  models}

We employ the approach suggested by \citet{m00} to calculate the
global structure of an accretion flow surrounding a Schwarzschild
black hole in the general relativistic frame. All the radiation
processes are included in the calculations of the global accretion
flow structure \citep*[see][for details and the references
therein]{m00,{2009ApJ...699..722Y}}. In this work, we include
outflows in the calculation of the ADAF structure, which have not
been considered by \citet{m00}, and the values of some parameters
adopted in this work are different from those in \citet{m00}.

The global structure of an accretion flow surrounding a black hole
with mass $M_{\rm bh}$ can be calculated, if some parameters
$\dot{m}$, $\alpha$, $\beta$ and $\delta$ are specified. The value
of the viscosity parameter $\alpha$ is still a controversial issue.
\citet{n96} assumed a fairly large value, $\alpha=1$, in order to
reproduce the maximum luminosities observed in low-state XRBs, while
\citet{e97} suggested a smaller value, $\alpha=0.25$, in their
spectral model calculations. Recently, the influence of viscosity
parameter $\alpha$ on the transition luminosity between the spectral
states of accreting X-ray binaries was explored with a wide range of
$\alpha$ from 0.1 to 0.9 \citep{q09}, which shows that the detected
transition luminosity between the spectral states of many galactic
accreting X-ray binaries can be fitted by tuning the value of
$\alpha$. In this work, we adopt two different values of $\alpha=
0.2$ and 0.5 in the calculations, respectively. The parameter
$\beta$, defined as $p_{\rm m}=B^2/8\pi=(1-\beta)p_{\rm tot}$,
($p_{\rm tot}=p_{\rm gas}+p_{\rm m}$), describes the magnetic field
strength of the accretion flow.  We assume $\beta= 0.8$ in all the
calculations. This parameter will mainly affect the radio spectrum
from the source, while {it has little effect on} the X-ray emission,
which we mostly focus on in this work. The parameter $\delta$
describes the fraction of the viscously dissipated energy directly
going into electrons in the accretion flow. It was pointed out that
a significant fraction of the viscously dissipated energy could go
into electrons by magnetic reconnection, if the magnetic fields in
the flow are strong \citep{bl97,bl00}. The value of $\delta$ is
still uncertain, and we adopt a conventional value of $\delta=0.1$
in all the calculations as that adopted in \citet{cao07}.

{In this work, the global two-temperature structures of the ADAF
models are derived numerically by solving the full relativistic
hydrodynamical equations under the modeling techniques suggested by
\citet{m00}. The main improvements of this solution compared with
the previous works \citep*[e.g.,][]{gp98} are that the advected
fraction of the dissipated energy $f$ in the energy equations is not
treated as a parameter and the detailed radiation processes
including synchrotron, comptonization, and bremsstrahlung are
calculated self-consistently.  Although \citet{m00} found that the
solution with $f=1$ is quite identical with the solution obtained by
fully solving the energy equations, especially in the innermost part
of the flow, {there is a difference in the surface density in the
outer region of the ADAF} \citep*[see Figure 4 in ][]{m00}, which is
the dominant {region contributing to} the thermal X-ray iron line
emissions we explored in this work. {In \citet{n99}, the advected
fraction of the dissipated energy $f$ is assumed to be a constant in
solving the global structures of the ADAFs \citep{qn99,gp98,e97},
and the continuum emissions from the ADAFs are computed using the
solved global structures of the ADAFs. Moreover, they have not
calculated the global structure of the ADAFs with winds, instead,
they simply adopted the radial velocity, angular velocity, and sound
speed of the flow derived from the global relativistic models of
\citet{gp98} under the assumption of a constant mass accretion rate.
The density  of the accretion flow is then calculated by applying
the continuity equation with a variable $r$-dependent mass accretion
rate \citep*[see][for the details]{qn99}.}

Given the values for parameters $\alpha$, $\beta$ and $\delta$, the
mass accretion rate $\dot{m}$ can be tuned to fit the observed X-ray
continuum spectrum when the outer radius of ADAF is fixed. In
low-mass BHXBs, where mass is transferred via Roche lobe overflow,
the circularization radius of the incoming stream is large ($R_{\rm
circ}\sim 10^{4}-10^{5}R_{\rm S}, R_{\rm S}=2GM/c^{2}$)\citep{e97}.
We assume the dimensionless outer radius of the ADAF surrounding the
black hole in GX 339-4 to be $r_{\rm out}=R_{\rm out}/R_{\rm
S}=5\times 10^{4}$ in our calculations. A constant mass accretion
rate $\dot{m}$ independent of radius is adopted for the pure ADAF
model.  The jet/outflow models were widely used to account for the
observed broadband spectra and the radio/X-ray correlations of GX
339-4  and other X-ray binaries in the low or off state, though the
origin of the outflows in the accretion flows is still unclear
\citep{c03,m01,m03,t08}. In this work, we employ a power-law
$r$-dependent mass accretion rate,
\begin{equation}
 \dot{m}=\dot{m}_{\rm out}(r/r_{\rm out})^p,
 \label{mdot}
 \end{equation}
to describe the outflows from the ADAF, where the power-law index
$p$ represents the strength of the outflows in the accretion flow,
and $\dot{m}_{\rm out}$ is the mass accretion rate at the outer
radius. It can be easily found that the pure ADAF model is a
specific case of ADAF with outflows model for $p=0$. The global
structure of the ADAF is calculated, and then the spectrum is
derived to reproduce the observed X-ray continuum by tuning the
values of
 $\dot{m}_{\rm out}$ and $p$.

\section{Calculations of line equivalent widths}

Based on the structure of the ADAF calculated as described in the
last section, we can compute the thermal X-ray iron line emission
from GX 339-4. The equivalent width of an emission line is defined
as \be {\rm EW}_{\rm line}=L_{\rm line}/L_{\nu} (\nu_{\rm line}),
\label{EW} \ee where $L_{\rm line}$ is the total luminosity of the
emission line, and $L_{\nu} (\nu_{\rm line})$ is the spectral
luminosity of the continuum at the energy of the line $\nu_{\rm
line}$. The total line luminosity $L_{\rm line}$ is available by
integrating the emission from the hot gas over the different shells
of the accretion flow, \be L_{\rm line}=\int^{r_{\rm out}}_{r_{\rm
in}}n_{\rm e}^{2}(r)~\epsilon_{\rm line}(T_{\rm e})~ 4\pi rHdr~
R_{\rm S}^{2}, \label{l_line} \ee where $n_{\rm e}$ is the electron
density at dimensionless radius $r=R/R_{\rm S}$, $r_{\rm in}= R_{\rm
in}/R_{\rm S}$ and $r_{\rm out}= R_{\rm out}/R_{\rm S}$ are the
inner and outer radii of the accretion flow. The density $n_{\rm e}$
and the vertical half thickness $H$ of the ADAF as functions of
radius are given by the derived global ADAF structure. The quantity
$\epsilon_{\rm line}$ is the line emissivity for unit density of
plasma at a certain radius, which is a function of the local
electron temperature.

Given the electron temperature of the plasma, we use the standard
software package Astrophysical Plasma Emission Code (APEC)
\citep{s01} to calculate the emissivity of the chosen line. We
assume a solar abundance and ionization equilibrium in the
calculations. {\citet{p00} found that departures from ionization
equilibrium were small in a model with $\alpha$=0.1, but they will
be more severe in the models with larger values of $\alpha$ adopted
here.} The APEC code includes collisional excitation, recombination
to excited levels and dielectronic satellite lines. It ignores
photoionization, which is a few percent effect at most \citep{n99}.
The spectral continuum luminosity $L_{\nu} (\nu_{\rm line})$ from
the accretion flow is calculated with the derived ADAF structure. We
compute the He-like and H-like line equivalent widths of iron,
EW$_{1}$ and EW$_{2}$, centered at energy $\sim6.7$ keV and
$\sim6.97$ keV, respectively. The solar metallicity $Z_{\odot}$ is
adopted in the calculations. The results can be scaled to other
metallicities $Z$ using  \be
 {\rm EW}_{1,2} (Z)=\frac{Z}{Z_{\odot}} {\rm EW}_{1,2} (Z_{\odot}).
 \ee

\section{Results}
The system parameters (and companion star) of GX 339-4 have not been
{directly measured}. \citet{h03} estimated a 1.7557 days orbital
period by analyzing the observed NIII (Bowen blend) lines and He II
wings, and derived a minimum mass of $5.8M_\odot$ for the black hole
in GX 339-4. The distance of GX 339-4 is estimated to be $d\gtrsim
6$ kpc from the optical {Na D line profiles}, and perhaps even to be
as large as 15 kpc \citep{h04}. \citet{z04} estimated the distance
to be $\gtrsim 7$ kpc based on Very Large Telescope (VLT)
observation of GX 339-4. We adopt $M=10M_{\odot}$ and $d=8$ kpc in
all the calculations, which are the same as those adopted in
\citet{t09}.

As discussed in \S 2, most of the disk parameters are fixed in the
calculations of the global structure of the ADAFs, except the
viscosity parameter $\alpha$, the parameter $p$ describing the
strength of the outflows, and the mass accretion rate at the outer
radius of the ADAF $\dot{m}_{\rm out}$. {The calculations are
carried out with different viscosity parameters, $\alpha=0.2$ and
$\alpha=0.5$, respectively.} The calculated global structures of
some ADAFs with different parameters are plotted in Figures 1 and 2.

\begin{figure}
\plotone{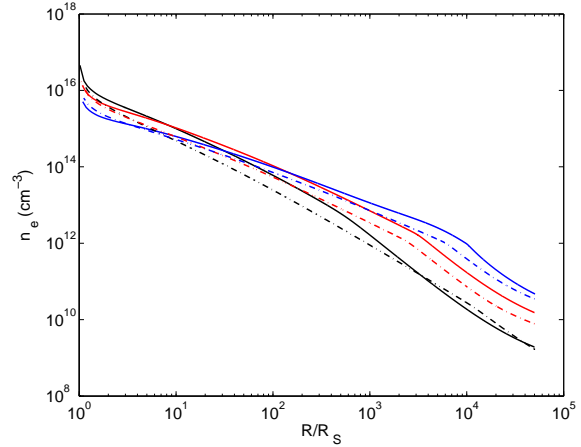} \caption{Electron density distributions of
the ADAF models with $\alpha=0.2$ and $\alpha=0.5$, respectively.
The black solid line corresponds to the pure ADAF model with
parameters $\alpha=0.2$, $\dot{m}=0.001$. The red and blue solid
lines correspond to the ADAF with outflows models with parameters
$\alpha=0.2$, $\dot{m}_{\rm out}=0.008$, $p=0.25$, and $\alpha=0.2$,
$\dot{m}_{\rm out}=0.025$, $p=0.45$, respectively. The black
dot-dashed line corresponds to the pure ADAF model with parameters
$\alpha=0.5$, $\dot{m}=0.001$. The red and blue dot-dashed lines
correspond to the ADAF with outflows models assuming parameters
$\alpha=0.5$, $\dot{m}_{\rm out}=0.01$, $p=0.25$, and $\alpha=0.5$,
$\dot{m}_{\rm out}=0.045$, $p=0.45$, respectively.}
\label{fig_density}
\end{figure}

\begin{figure}
\plotone{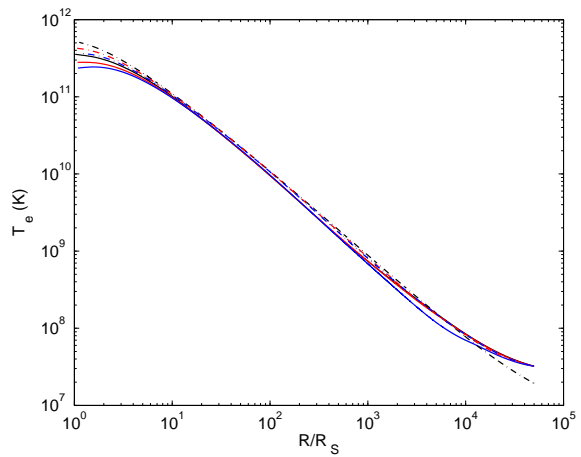} \caption{Same as Figure 1, but for the
electron temperature distributions.} \label{fig_temp}
\end{figure}

Based on the derived ADAF structure, the continuum spectrum can be
calculated \citep*[see][for the details]{cao05,cao07}. {The value of
the mass accretion rate at the outer radius of the ADAF
$\dot{m}_{\rm out}$ and the outflow strength parameter $p$ are tuned
in order to fit the observed X-ray continuum spectrum between 0.4-11
keV\citep{g03}.} It is found that the values of $\dot{m}_{\rm out}$
and $p$ are degenerate, i.e., the X-ray continuum spectrum can be
fairly well reproduced with different sets of the parameters
$\dot{m}_{\rm out}$ and $p$ (see Figure \ref{fig_spectr}).

\begin{figure}
\plotone{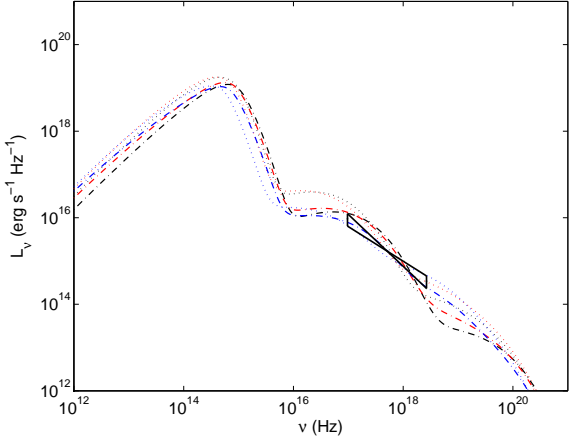} \caption{Spectra of  ADAF models with
$\alpha=0.2$ and $\alpha=0.5$, respectively. The black dotted line
corresponds to the pure ADAF model with parameters $\alpha=0.2$,
$\dot{m}=0.001$. The red and blue dotted lines correspond to the
ADAF with outflows models with parameters $\alpha=0.2$,
$\dot{m}_{\rm out}=0.008$, $p=0.25$, and $\alpha=0.2$, $\dot{m}_{\rm
out}=0.025$, $p=0.45$, respectively. The black dot-dashed line
corresponds to the pure ADAF model with parameters $\alpha=0.5$,
$\dot{m}=0.001$. The red and blue dot-dashed lines correspond to the
ADAF with outflows models assuming parameters $\alpha=0.5$,
$\dot{m}_{\rm out}=0.01$, $p=0.25$, and $\alpha=0.5$, $\dot{m}_{\rm
out}=0.045$, $p=0.45$, respectively. The black tie represents the
observed 0.4-11 keV X-ray spectrum from \citet{g03}.}
\label{fig_spectr}
\end{figure}

\begin{deluxetable}{ccccc}
\tabletypesize{\scriptsize} \tablecaption{Equivalent widths EW$_{1}
(Z_{\odot})$ and EW$_{2} (Z_{\odot})$ in units of eV of the thermal
helium-like and hydrogen-like  X-ray iron lines from the ADAFs}
%\tablecaption{from $1.^{\prime\prime}5$ of Sgr A* for five models}\\
\tablewidth{0pt} \tablehead{ \colhead{$\alpha$} &\colhead {
$\dot{m}_{\rm out}$}&\colhead{ $p$}&\colhead{
EW$_{1}$(eV)}&\colhead{ EW$_{2}$(eV)} } \startdata
 0.2 & ~ 0.001 &~0.~~  & ~~~~0.7 &~~~~0.7
\\
0.2 & ~ 0.008& ~  0.25   & ~ 28.9 & ~  24.3
\\
0.2 & ~ 0.01~~ & ~ 0.3~   & ~ 54.5 & ~  41.1
\\
0.2 & ~ 0.015& ~  0.35   & ~103.9 & ~  69.0
\\
0.2 & ~ 0.02~~ & ~  0.4~  & ~196.2 & ~110.1
\\
0.2 & ~ 0.025& ~  0.45   & ~368.0 & ~169.2
\\\\
0.5  & ~ 0.001 & ~ 0.~~~ & ~~~ 0.4 & ~~~  0.5
\\
0.5 & ~ 0.01~~ & ~ 0.25 & ~~~ 7.4 & ~~~  6.5
\\
0.5 & ~  0.015 & ~0.3~ & ~ 17.0 & ~  13.6
\\
0.5 & ~  0.025 & ~0.35 & ~ 38.8 & ~ 27.5
\\
0.5 & ~  0.035 & ~0.4~ & ~  77.2 & ~ 48.2
\\
0.5 & ~  0.045  & ~0.45  & ~  141.0 & ~ 77.1 \\
\enddata
%\tablecomments{These lines are from the accretion flow. }

\end{deluxetable}

Assuming a solar metallicity for the plasmas in the ADAF, we
calculate the equivalent widths of the thermal He-like and H-like
iron lines emitted from the accretion flows as described in \S3. The
results are listed in Table 1, where EW$_{1}(Z_{\odot})$ and
EW$_{2}(Z_{\odot})$ represent the equivalent widths of the thermal
He-like and H-like iron lines, respectively.  The adopted parameters
{are also shown} in Table 1. The total equivalent widths of the
He-like and H-like iron lines emitted from the ADAFs surrounding the
black hole in GX 339-4 are plotted in Figure 4 as a function of the
outer mass accretion rate for the models with different strength of
the outflow. {The results show that the predicted EWs of the He-like
and H-like iron lines from the pure ADAFs are too small to be
observed. For the ADAFs with outflows, the EWs of the thermal iron
lines significantly increase with the strength of the outflow.}

\begin{figure}
\plotone{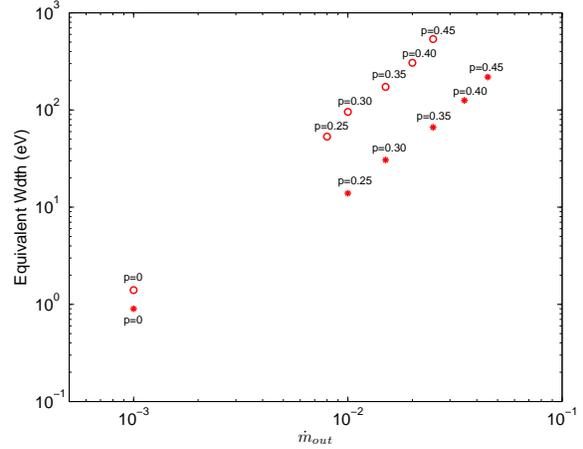} \caption{Total equivalent widths  of the
He-like and H-like iron lines from the ADAFs as a function of the
outer mass accretion rate. Circles and stars represent the data for
models with $\alpha=0.2$ and $\alpha=0.5$, respectively. The
power-law indexes, $p$, in Equation (\ref{mdot}) are indicated in
the figure for different ADAF models, while $p=0$ corresponds to the
pure ADAF models. } \label{fig_ew}
\end{figure}

\section{Discussion}

In this paper we have explored the thermal X-ray line emission from
the accretion flow surrounding the black hole in the galactic binary
GX 339-4.  We construct the ADAF models for the observed X-ray
spectrum of GX 339-4 with different parameters including the
viscosity parameter $\alpha$, the mass accretion rate $\dot{m}$ and
the outflow strength $p$. We compute the equivalent widths of the
helium-like and hydrogen-like thermal iron lines from the accretion
flows.

From Table 1 and Figure 4, we can see that the thermal X-ray line
emission from the ADAFs without outflows is always much lower than
emission from the ADAFs with outflows. Moreover, the equivalent
widths of the He-like and H-like thermal iron lines significantly
increase with the mass accretion rate $\dot{m}_{\rm out}$ at the
outer radius of the ADAF and the outflow strength parameter $p$. In
fact, these two parameters are not independent, i.e., the outflow
strength parameter $p$ is required to increase with the mass
accretion rate $\dot{m}_{\rm out}$ at the outer radius to reproduce
the observed X-ray continuum spectrum. The sensitive dependence of
EW$_{1}$ and EW$_{2}$ on the two parameters can be explained by the
very different density profiles (see Figure 1), because the
temperature profiles of the plasmas are quite similar for ADAFs with
different parameters adopted \citep*[see Figure 2, and][]{ny95}. As
showed in Equation (\ref{EW}), the equivalent width of a thermal
line is the ratio of the line luminosity and the spectral luminosity
at the energy of the line. All our derived spectra are required to
fit the observed X-ray continuum at $\sim 6.83$ keV. Thus, the
equivalent widths of thermal iron lines will mainly depend on the
line luminosities. From Equation (\ref{l_line}), the line luminosity
emitted from a certain annulus of the accretion flow is determined
by the density and the line emissivity, which is a function of the
electron temperature. As the temperature profiles are quite similar
for the models with different parameters, the line luminosity is
mostly determined by the density distribution of the gas in the
accretion flow. For the models with larger $p$, i.e., stronger
outflows, the density profiles are flatter, which means that much
more gases are blown away from the accretion flow. This leads to
higher electron densities in the outer region of the accretion flow
($r\gtrsim 10^{3}$) with the electron temperature $\lesssim10^8$ K,
where the thermal He-like and H-like iron lines emitted. Therefore,
the line luminosity and the equivalent widths of the thermal iron
lines are higher in these models.

{The change of the equivalent widths of the thermal iron lines
emitted from the ADAFs with the value of the viscosity parameter
$\alpha$ is explored in this work. Comparing the two groups of the
line equivalent widths for different values of viscosity,
$\alpha=0.2$ and $\alpha=0.5$, we find that the EWs of the lines
from the ADAFs with $\alpha=0.2$ are always larger than those from
the ADAFs with $\alpha=0.5$ if the same value of $p$, i.e., strength
of the outflows, is adopted. This is because the larger the
viscosity is, the quicker the gas flows in, which leads to smaller
densities of the gas in the outer region of the ADAF (see Figure
1).}

{We theoretically conclude in this work that the galactic X-ray
binary GX 339-4 may produce  X-ray iron lines with very large
equivalent width in the quiescence. For reasonable choice of model
parameters, e.g., $\alpha=0.2$, $\dot{m}_{\rm out}=0.25$ and
$p=0.45$, the total equivalent width of the He-like and H-like iron
lines from the ADAFs with moderately strong outflows reaches to
$\gtrsim 500$ eV, when the 0.4-11 keV X-ray flux is about $3.8\times
10^{-13}{\rm erg~cm^{-2}~s^{-1}}$. If these strong thermal lines are
really observed, it may give a clue to the accretion mode of the
source in the off state and provide strong evidence that the
outflows/winds exist in GX 339-4. In fact, \citet{f01} showed that a
prominent thermal X-ray iron line, with the equivalent width of
$595^{+248}_{-172}$ eV and the line energy centered at
$6.83^{+0.16}_{-0.16}$ keV, is present in the RXTE (PCA) observation
of GX 339-4 in 1999 when the source is in the off state with the
integrated 3-20 keV flux of $7\times 10^{-12}{\rm
erg~cm^{-2}~s^{-1}}$.  However, we need to bear in mind that the
observation with RXTE may include possible contamination from the
Galactic ridge or {other sources}. Observations with ASCA, Suzaku
and Chandra have detected an iron line emission at 6.7 keV and
suggested it to be emitted {by the diffuse} hot plasma and/or
produced by point sources \citep*[see, e.g.,][]{1986PASJ...38..121K,
2002A&A...382.1052T, r09}. Thus, we {select the Chandra} observation
\citep{g03} as the quiescence flux in our calculations.}

{Due to the different sources we explored and different parameters
we used in the calculations, it is not easy to directly compare the
equivalent widths of thermal iron lines from GX 339-4 derived in
this work with those from V404 Cyg predicted by \citet{n99}.
Therefore, we recalculate the equivalent widths of the thermal iron
lines emitted from V404 Cyg in quiescence with our ADAF model for
the same model parameters as those used in \citet{n99}. The
comparison of the {results is shown} in Table 2. The value of
parameter $\beta$ in \citet{n99} (defined as $\beta=p_{\rm
gas}/p_{\rm mag}$) has been translated to the value with the
definition in this work. The equivalent width of the He-like iron
line EW$_{1}$ of \citet{n99} is the sum of the several He-like lines
with the energy between 6.65 keV and 6.75 KeV which can be compared
with the result of this work, because we have already combined the
equivalent widths of He-like lines near the energy of 6.7 keV in
bins of 100 eV width which we believe to be more comparable with the
observation. Moreover, we note that the equivalent width of the
H-like iron line of \citet{n99} only includes one line Fe XXVI
$\lambda$1.780 while the data of this work is also the sum of the
H-like lines in bins of 100 eV width, i.e. with the energy between
6.92 keV and 7.02 keV. {The emission in the 100 eV bin is larger
than that of the line calculated by \citet{n99}.} The calculated
X-ray continuum derived from the approach of this work is a bit
larger than the ASCA observation of V404 Cyg
\citep{1997ApJ...482..448N} if the same mass accretion rate
$\dot{m}_{\rm out}$ as \citet{n99} is adopted. Using the same model
parameters $\delta,\alpha,\beta$ and the strength of the wind $p$ as
\citet{n99}, our calculation shows that the best fitted mass
accretion rate is $\dot{m}_{\rm out}=0.004$ (the results are listed
in Table 2). The discrepancy of the results between this work and
\citet{n99} is significant, which is mainly due to the difference of
ADAF structure calculations.}

\begin{deluxetable}{cccccccc} \tabletypesize{\scriptsize}
\tablecaption{Comparison of the Equivalent widths EW$_{1}
(Z_{\odot})$ and EW$_{2} (Z_{\odot})$ of the thermal helium-like and
hydrogen-like  X-ray iron lines from V404 Cyg calculated by the
solutions in this work and in \citet{n99}} \tablewidth{0pt}
\tablehead{\colhead{Data}&\colhead{$\delta$}&\colhead{$\alpha$}
&\colhead{$\beta$}&\colhead{ $p$}&\colhead { $\dot{m}_{\rm
out}$}&\colhead{ EW$_{1}$(eV)}&\colhead{ EW$_{2}$(eV)} } \startdata
NY99$^{(1)}$ &~0.3 & ~0.1   &~0.91 &~0.4 & ~ 0.006& ~230 &
~79$^{(2)}$
\\
this work &~0.3 & ~0.1   &~0.91 &~0.4 & ~ 0.006&~182 &~113~~~~
\\
this work &~0.3 & ~0.1   &~0.91 &~0.4 & ~ 0.004&~122 &~79~~~~
\\
\enddata
\tablecomments{(1) NY99 represents the data from \citet{n99}. (2)
EW$_{2}$ of \citet{n99} includes only  one line Fe XXVI
$\lambda$1.780 while the data of this work is the sum of the H-like
iron lines in bins of 100 eV width, i.e. with the energy between
6.92 keV and 7.02 keV. }

\end{deluxetable}

\acknowledgments  We thank the anonymous referee for very helpful
suggestions. This work is supported by the NSFC (grants 10778621,
10703003, and 11078014).

{}

%\end{document}


%\clearpage

\end{document}